\pdfoutput=1

\documentclass[aip,jap,reprint]{revtex4-1}

\usepackage{graphicx}
\usepackage{amsmath}

\newcommand{\REM}[1]{}

\newcommand{\fig}[1]{Fig.~\ref{#1}}
\newcommand{\eqn}[1]{Eqn.~\ref{#1}}
\newcommand{\Fig}[1]{Fig.~\ref{#1}}

\newcommand{\ten}[1]{\ensuremath{\times 10^{#1}}}
\newcommand{\unit}[1]{\ensuremath{\mathrm{\: #1}}}

\begin{document}

\title{Shape oscillations of a charged diamagnetically-levitated droplet}

\author{R.J.A. Hill}
\author{L. Eaves}
\affiliation{School of Physics and Astronomy, University of Nottingham, Nottingham NG7 2RD, UK}
\email{richard.hill@nottingham.ac.uk}

\date{\today}

\begin{abstract}
We investigate the effect of electrical charge on the normal mode frequencies of electrically-charged diamagnetically levitated water droplets with radii 4.5-7.5~mm using diamagnetic levitation. This technique allows us to levitate almost spherical droplets and therefore to directly compare the measured vibrational frequencies of the first seven modes of the charged droplet with theoretical values calculated by Lord Rayleigh, for which we find good agreement.    
\end{abstract}

\maketitle
%\section{Introduction}
Understanding how electrically-charged liquid droplets vibrate, distort and shatter is important in high technology areas such as ink jet printing and electrospray ionization for mass-spectrometry of biomolecules. The quiescent shapes and dynamics of the charged drop influence the echos returned from rainfall radar, affect raindrop breakup, and possibly lightning activity \cite{braziersmith71,bhalwankar09,beard10}. Based on some recent reports \cite{mchale07,mchale09}, we anticipate that the effects of charge on the vibrations of `liquid marbles', microliter powder-coated droplets with applications in microfluidics \cite{aussillous01}, may also become an interesting topic for future study. The charged liquid droplet is the basis of Bohr and Wheeler's liquid drop model of the atomic nucleus and one might hope to gain new insights into the nucleus and its fission by studying the dynamics of this classical system. 

The starting point for understanding these areas is Lord Rayleigh's calculation of the frequency of the normal modes of an electrically-charged droplet, made in 1882 \cite{rayleigh1882}. However, it has been surprisingly difficult to verify experimentally Rayleigh's result directly, owing to the requirement that the droplet be spherical and the measurement taken without contacting the liquid. 

Brazier-Smith et al. measured the oscillations of a single 1.05~mm radius water droplet suspended on a wire, as a function of voltage \cite{braziersmith71}, and Saunders and Wang verified this result for droplet radii up to 2.8 mm using air-flow to suspend the droplets \cite{saunders74}. They obtained good agreement with theory for the fundamental mode, but were unable to measure the oscillations of higher order modes. Trinh et al. obtained measurements of the first three modes using a hybrid acoustic and electrostatic levitation technique, but were unable to compare their results directly with Rayleigh's theory, owing to the non-spherical equilibrium shape of the droplet levitated in this way \cite{trinh96}. The equilibrium shape of an acoustically-levitated or air-flow levitated droplet is non-spherical because the gravitational force on the whole droplet is balanced only at the liquid's surface. Hence, the droplet becomes distorted under its own weight. The frequency of the fundamental mode of small, 1-50 $\mu$ m diameter, nearly spherical water droplets, levitated electrostatically, has been shown to agree with Rayleigh's theory \cite{richardson89, duft02, duft03, giglio08}, but measurements of the frequencies of higher order modes have not been made using this technique.

Here, we use diamagnetic levitation to study the normal mode oscillations of an electrically-charged water droplet, with radii $a$ between 4.5 and 7.5~mm, and surface electric charge of $10^{-9}\unit{C} \sim 10^{10}e$. Previous experiments have used diamagnetic levitation to study the dynamics of uncharged droplets \cite{beaugnon91jphys,beaugnon91nat,beaugnon01,sueda07,hill08,hill10,hill12} and formation of liquid films \cite{katsuki08,tanimoto08}. In contrast to acoustic, electrostatic or air flow methods, diamagnetic levitation balances the force of gravity at the molecular level throughout the body of the levitating fluid, effectively allowing one to study fluid dynamics in a nearly zero gravity environment. With this technique, one can levitate large mm to cm-scale droplets that are nearly spherical at rest. The large size of the droplets allows us to make accurate measurements of the frequencies of the normal modes using a relatively simple technique, and because the drops are spherical, we can compare the measurements directly with Rayleigh's theory.  We observe the vibrations of the fundamental and higher order modes, up to the first seven modes and find good agreement with Lord Rayleigh's theory for all these modes. Whitaker et al. and Vicente et al. \cite{whitaker98shapeosc, whitaker98charged, vicente02} have measured the resonant frequencies of the first 14 modes of diamagnetically levitated helium droplets, 1-3~mm in radius, with charge $\sim 10^3-10^4$ times smaller than in our experiments. However, in their experiments, the small charge was applied as part of a technique to excite vibrations by an electric field, and was too small to give a measurable charge-induced shift in frequency.
% 629

\begin{figure}
 \includegraphics[width=83mm]{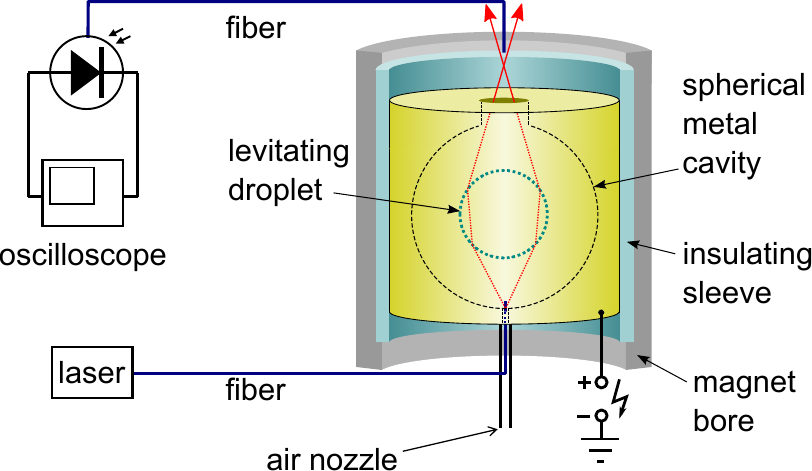}
 \caption{Cutaway drawing of a water droplet levitating within a spherical brass cavity inside a plastic sleeve within the 50mm-diameter magnet bore. See text.}
\label{f1}
\end{figure}

%\section{Method} \label{experimental_details}
We use a superconducting solenoid magnet with a vertical, room-temperature, 50~mm diameter bore to levitate the water droplets. A water droplet levitates approximately 80~mm above the geometric center of our solenoid, where the diamagnetic force, proportional to $B\nabla B$, is equal in magnitude to, and opposite in direction to, the gravitational force on (i.e. the weight of) the droplet \cite{berry97,simon00}. The magnetic field is $B\approx 12\unit{T}$ and the vertical field gradient is $\partial B/\partial z \approx 120\unit{Tm^{-1}}$ at the levitation point. We fine-tune the current in the solenoid until the equilibrium shape of the levitating droplet is as close to spherical as possible. This is achieved by reducing the quadrupole component of the magnetogravitational potential `trap' to zero; details of this procedure are given in Ref.~\onlinecite{hill10}. Droplets of the sizes we use here are spherical to a good approximation. We measure the ratio of the equatorial (horizontal) diameter to the polar (vertical) diameter to be $1.00\pm0.02$.
 
\fig{f1} shows a schematic diagram of the water droplet levitating in the vertical magnet bore. The droplet levitates within a spherical metal cavity, which is electrically insulated from the magnet bore by a plastic sleeve. A voltage is applied between the cavity and earth by using a high voltage power supply. 

A 1~mm-diameter nozzle passes through the hole at the base of the cavity, and is directed at the center of the underside of the droplet, as shown in \fig{f1}. It is connected to a rubber bulb outside the magnet by a tube. When the bulb is struck on a hard surface, the resulting pulse of air from the nozzle excites several shape oscillation modes simultaneously, with amplitude less than $5\%$ of the droplet radius.

An optical fiber, inserted through the air nozzle, directs light from a HeNe laser at the droplet. The drop acts as a lens, focusing the light onto the aperture of a second fiber, through a hole at the top of the spherical cavity, as shown in the figure. This fiber transmits the light to a photodiode outside the magnet. The emf of the photodiode is recorded by a digital storage oscilloscope. Since the focal length of the drop depends on its shape, the intensity of the laser light falling on the photodiode oscillates as the drop vibrates. Other groups have employed variations of this optical technique for measuring the shape oscillations (e.g. Refs.~\onlinecite{whitaker98charged,duft02}).  

The temperature of the water was brought to $16^{\circ}$C in a water bath before it was injected into the potential trap, to match the ambient temperature in the magnet bore. At this temperature, the density of the liquid is $\rho = 999 \unit{kg m^{-3}}$, its surface tension is $T = 73.3 \unit{mNm^{-1}}$ and its kinematic viscosity is $\nu = 1.11 \ten{-6}\unit{m^2s^{-1}}$ \cite{lemmon}. The liquid was injected into the trap using a glass pipette. The volume of liquid injected was determined to better than $1\%$ uncertainty from the difference in the weight of the pipette before and after injection. The liquid was drawn out of the bore using a paper towel by capillary action after the experiment. By measuring the difference in weight between the wet and dry paper, we obtained a second measurement of the droplet volume. Using this simple technique, we were able to determine that the mass loss through evaporation of the drop, during the measurement period (approximately 30-60 minutes), was less than $2\%$ for all the droplets we measured, except for the smallest ($4.5$~mm radius droplet), where the loss was $3\%$.

The procedure for charging the droplet is as follows. A thin wire, connected to earth, is brought into contact with the droplet through the hole in the top of the cavity. A voltage is applied between earth and the cavity enclosure, charging the droplet capacitively. Since the cavity is spherical, the charge on the droplet, as a function of voltage, is given by
% 642

\begin{equation}
Q = 4\pi \epsilon_0 V \left (\frac{1}{a} - \frac{1}{R} \right )^{-1},
\end{equation}
where $R$ is the internal radius of the cavity, $a$ is the radius of the spherical droplet at rest, and $V$ is the voltage between the wire and the cavity.

The wire is slowly removed from the droplet and then the voltage on the cavity is returned to ground. The volume of liquid withdrawn with the fine wire is negligible. Other methods have also been used to charge levitating droplets (e.g. Ref.~\onlinecite{hilger07}). The charge on the droplet leaks away with a time constant of several hours, allowing ample time to measure the normal mode frequencies. The maximum voltage applied between cavity and droplet was 3kV. Larger voltages caused the mechanical equilibrium between diamagnetic, electric and gravitational forces to become unstable, causing the droplet to escape the magnetogravitational potential `trap' and hit the walls of the cavity.
% 136

\begin{figure}
\includegraphics[width=83mm]{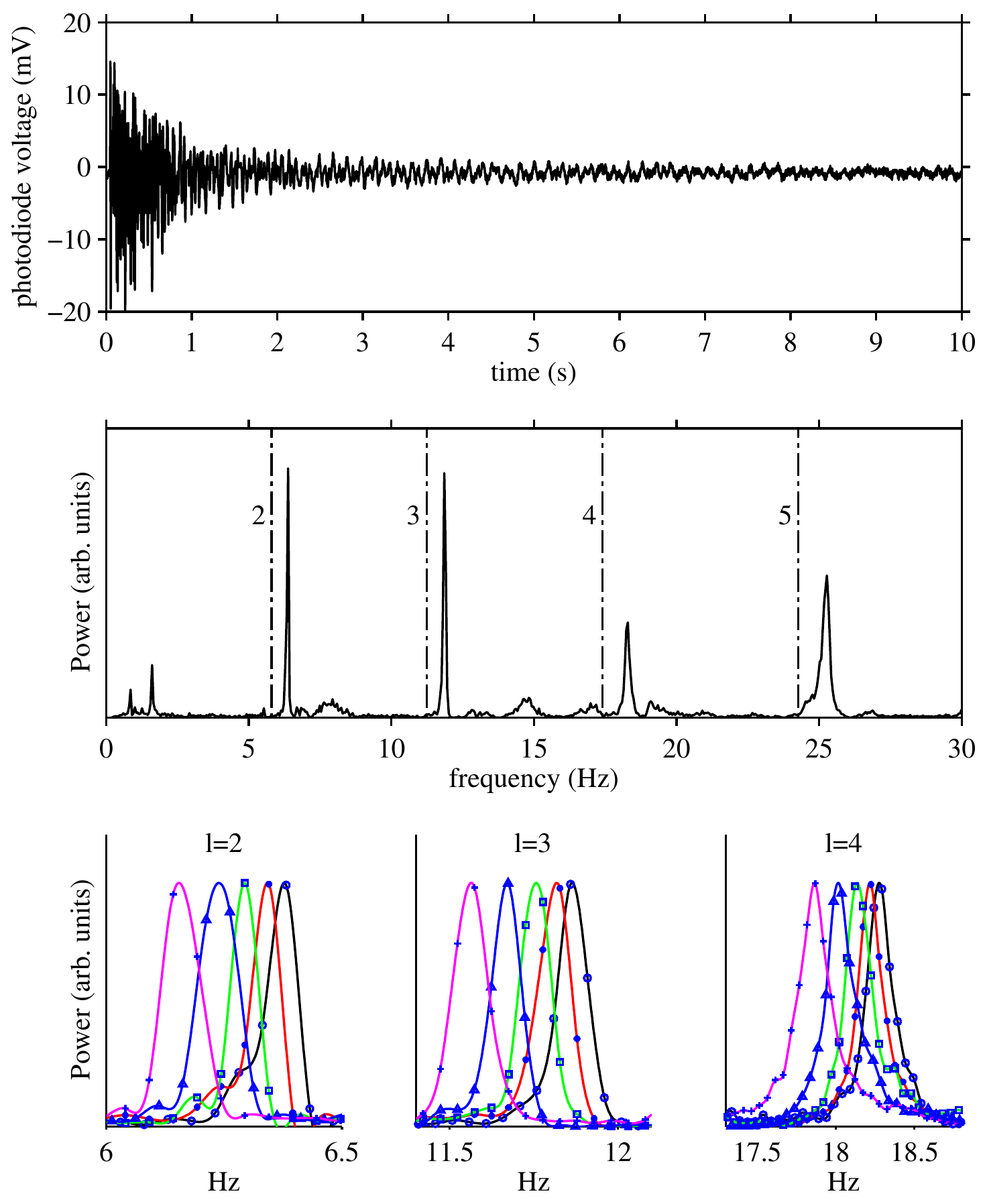}
\caption
{Top: Oscillations in laser light intensity falling on the photodiode reveal the oscillations of a water droplet with radius 7.5mm. Center: The peaks in the power spectrum of the oscillations show clearly the $l=2-5$ resonant frequencies of the droplet. The measured resonant frequencies are slightly higher than the frequencies predicted by \eqn{e1}, indicated by dashed lines, owing to the effect of the magnetogravitational potential trap. Bottom: The peaks shift to lower frequency with increasing charge. Black (open circle) uncharged; red (filled circle) 1.4nC; green (square) 2.2nC; blue (triangle) 2.9nC; pink (cross) 3.6nC.}
\label{f2}
\end{figure}

%\section{Results}
\Fig{f2} top shows the oscillations in the photodiode signal developed by the light refracted through an $a=7.5\unit{mm}$ droplet; they decay exponentially as $\exp(-\lambda t)$ with a time constant $1/\lambda = 0.1-10\unit{s}$, dependent on $a$ and the oscillation mode. The center panel of \Fig{f2} shows the power spectrum of these oscillations. Several peaks are evident in this spectrum, corresponding to the $l=2$ to 5 Rayleigh modes. The peak near 2~Hz is due to a small oscillation of amplitude $\sim 1\unit{mm}$ of the droplet's center of mass of about the equilibrium levitation position. 

The lower panels show the power spectrum in three narrow frequency windows, centered on the $l=2$, 3 and 4 resonances. The resonances move to lower frequency with increasing charge $Q$. 

For small amplitude oscillations, the resonant frequencies are expected to be $\omega^2 = \omega_0^2 + \omega_C^2$, where  
\begin{equation}
  \omega_0^2 = \frac{Tl(l-1)(l+2)}{\rho a^3}
\label{e1}
\end{equation}
is the resonant frequency of the $l$th mode of an uncharged droplet with surface tension $T$ and density $\rho$, and 
\begin{equation}
  \omega_C^2 = - \frac{Q^2 l(l-1)}{16 \pi^2 \epsilon_0 \rho a^6}
\label{e2}
\end{equation}
is the shift in frequency owing to the charge $Q$ on the surface \cite{rayleigh1882}. These equations assume that the viscosity is zero. We justify this assumption shortly. Since the Coulomb forces between the charges at the surface oppose the cohesive force of surface tension, increasing the surface charge density reduces progressively the frequencies of the normal modes until the critical charge $Q_R^2 = 64 \pi^2 \epsilon_0 a^3 T$ is reached, known as the Rayleigh limit. At this point the frequency of the fundamental mode ($l=2$) is reduced to zero, and the droplet becomes unstable to break up under the action of the Coulomb force \cite{richardson89, duft02, duft03, giglio08}. Here, the largest voltage that we apply charges the droplets to approximately $Q_R/3$, which is approximately $1\times10^{10}e$ for $a=4$~mm, and $3\times10^{10}e$ for $a=8$~mm.

The dotted vertical lines in \fig{f2} show the resonant frequencies of an uncharged droplet, as given by \eqn{e1}. The measured frequency of each resonance is slightly higher than we expect from \eqn{e1}. This is due to the effect of the magnetogravitational potential trap, which acts as an additional cohesive force on the droplet. The magnetogravitational potential trap, within which the droplet levtitates, has the form $U(r,\theta) = c_0(r) + c_2(r) P_2(\cos \theta) + c_3(r)P_3(\cos \theta)$, where $r$ and $\theta$ are spherical polar coordinates centered on the levitation point ($\theta$ is the inclination) and $P_n$ are Legendre functions. In these experiments, the quadrupole component $c_2 P_2$ has been reduced to zero by adjusting the current in the solenoid. There is no effect to first order in oscillation amplitude and $c_3$ on the frequencies. The spherical component of $U$ adds a term $c_0'(a) l/a$ to \eqn{e1}, where $c_0'(a) = \partial c_0/\partial r$. This effect is discussed in Ref.~\onlinecite{hill10,hill12}; see also Ref.~\onlinecite{vicente02}.
% 470

\begin{figure}
\includegraphics[width=83mm]{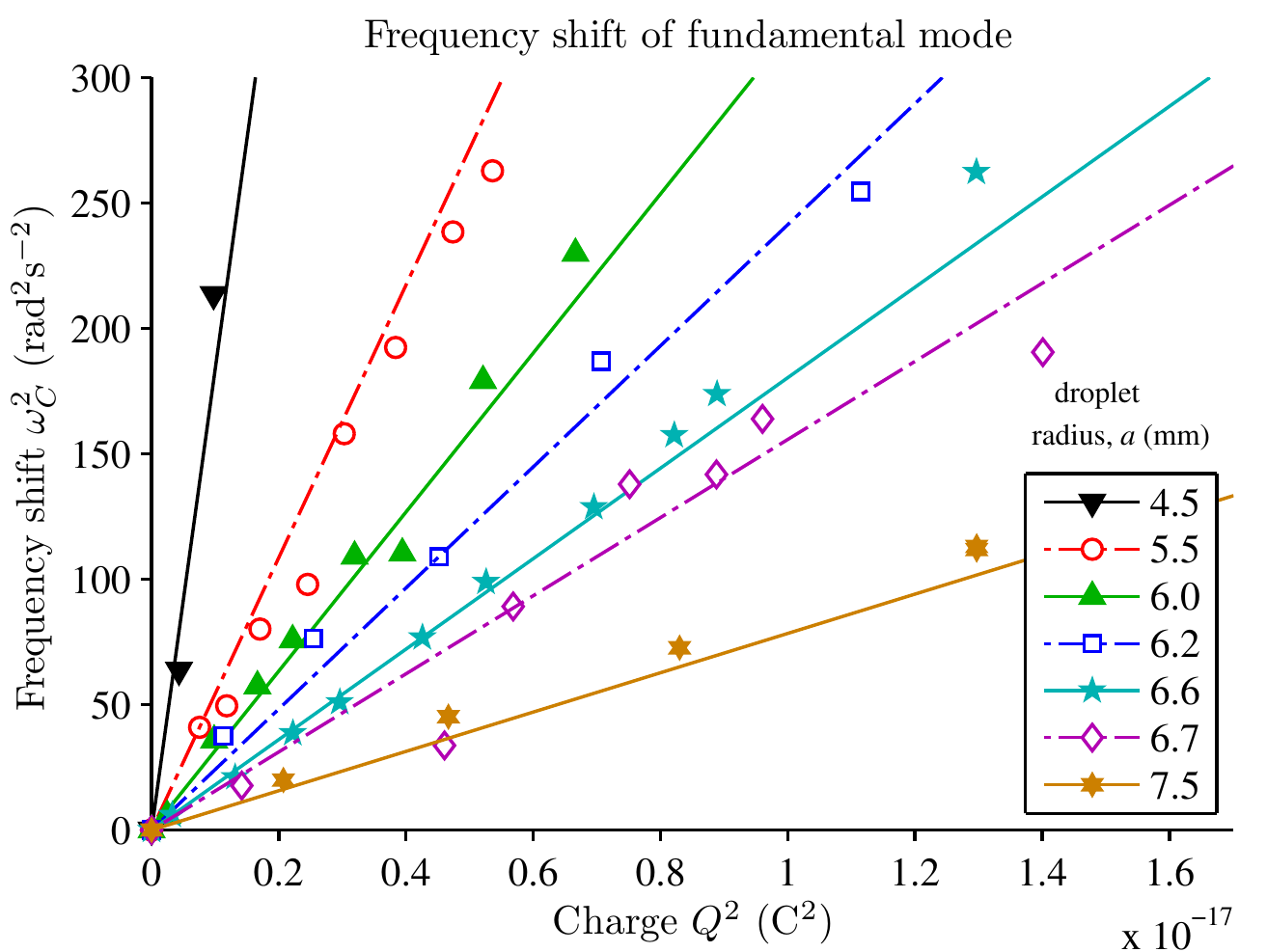}
\caption
{Shift in frequency $\omega_C^2 = \omega(0)^2 - \omega(Q)^2$ as function of charge $Q$ for the $l=2$ resonances. 
}
\label{f3}
\end{figure}

\begin{figure}
\includegraphics[width=83mm]{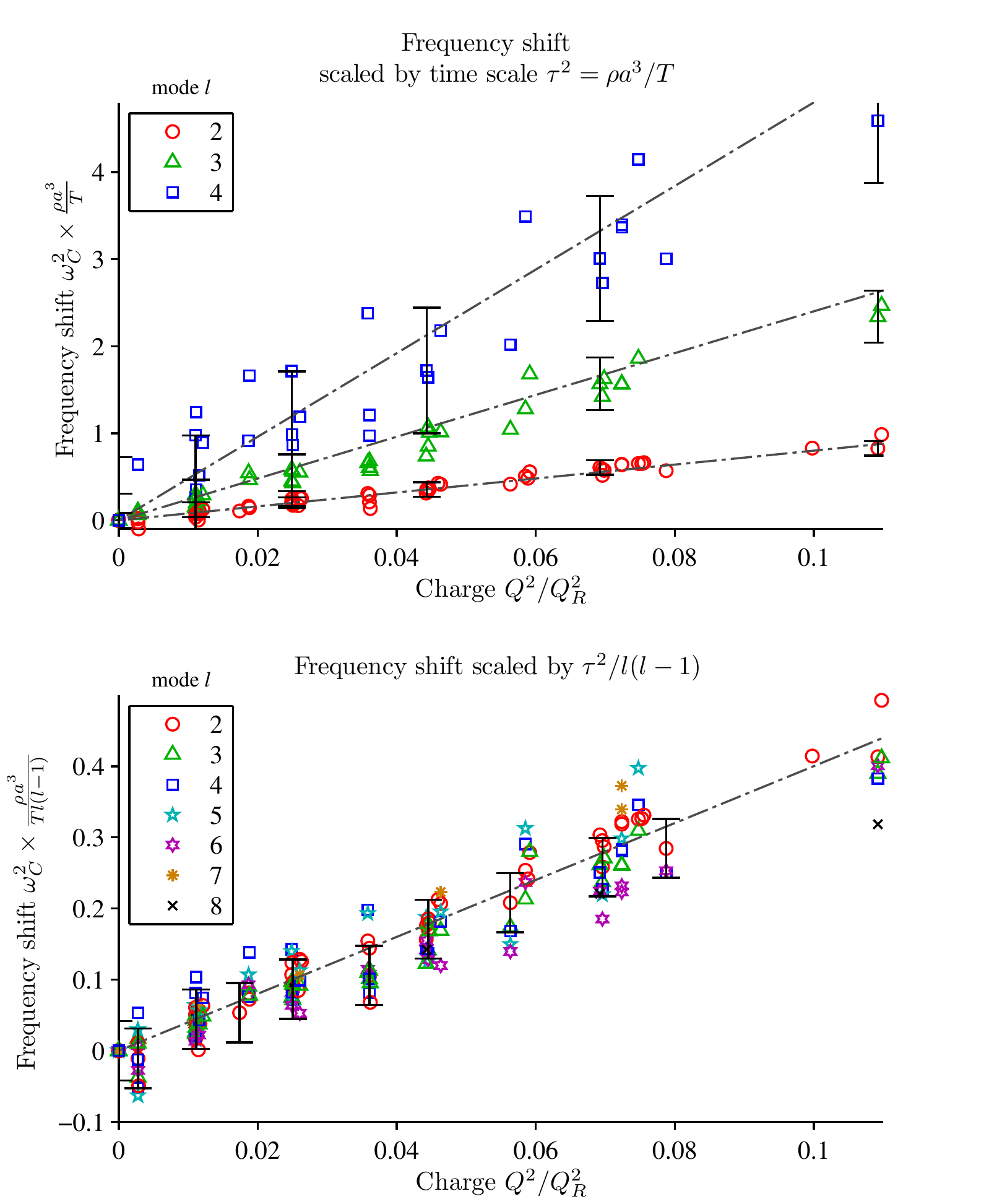}
\caption
{Top: Fan plot showing the shift in frequency $\omega_C^2 = \omega(0)^2 - \omega(Q)^2$ with charge $Q$ for the $l=2-4$ resonances. The frequency has been scaled by the time scale $\tau = (\rho a^3/T)^{1/2}$ and the charge by $Q_R^2 = 64 \pi^2 \epsilon_0 a^3 T$. Bottom) when the frequency shift $\omega_C^2$ is further scaled by $l(l-1)$, the data for all the modes collapse onto the same line. This panel includes the measured peak positions of modes $l>4$. Only selected data points shown with error bars, for clarity. 
}
\label{f4}
\end{figure}

\Fig{f3} shows the measured shift in frequency $\omega_C^2 = \omega(0)^2 - \omega(Q)^2$ as function of charge, for the $l=2$ resonances, and \fig{f4}, top, shows the shift in frequency $\omega_C^2$ for the $l=2-4$ resonances. In the latter figure, we have plotted the dimensionless frequency versus the dimensionless charge by scaling the frequency by the time scale $\tau = (\rho a^3/T)^{1/2}$ and the charge by $Q_R$. When the data are scaled this way, the resonances from droplets with different radii, ranging from $a=4.5$ to $a=7.5$~mm collapse onto the straight lines $\omega_C^2 \tau^2 = 4 l(l-1) Q^2/Q_R^2$ in agreement with \eqn{e2}. 

When the frequency shift $\omega_C^2$ is further scaled by $l(l-1)$, as in the bottom panel of \fig{f4}, the data for all the modes collapse onto the same line. This panel includes the measured peak positions of modes up to $l=8$, in addition to the data shown in the top panel. 

The most significant contribution to the error in $\omega_C^2$ comes from the uncertainty in the position of the peaks in the power spectrum. Although the uncertainty in the frequency of a particular resonance is approximately constant $\delta \omega \approx 0.03 \omega$, the uncertainty in the frequency \emph{difference} scales with $\omega$: $\delta \omega_C^2 = 2(\delta \omega/\omega) (\omega(Q)^2+\omega(0)^2) \approx 4 (\delta \omega/\omega) \omega_0^2$. Hence the uncertainty in $\omega_C^2$ is approximately proportional to $l(l-1)(l+2)$, as indicated by the error bars on \fig{f4}.

The uncertainty in the volume of the droplet comes primarily from the evaporative loss, which is less than $2\%$ for all the droplets we measured, except for the smallest ($a=4.5$~mm), where the loss was $3\%$. We estimate the error in $Q$ to be approximately $3\%$ from the combined uncertainties in $a$ and voltage.

Although a strong magnetic field is expected to damp the motion of a conducting liquid through the field via the Lorentz force\cite{priede11}, the conductivity of the distilled water used here is too small for any effect on the oscillation frequencies, or damping of the oscillations to be observed. 

The time constant $\lambda$ of the decay of the oscillations depends on the charge on the droplet. For example, for an $a=6$~mm droplet, the $l=2$ mode of the uncharged droplet has a time constant $\lambda = 0.16 \pm 0.02\unit{s^{-1}}$ in good agreement with the theory $\lambda \approx \nu a^{-2}(l-1)(2l+1)$ for low viscosity liquids \cite{tang74}. When we charge the droplet to $Q_R/3$, the oscillations decay more quickly, with a time constant $\lambda = 0.26 \pm 0.02\unit{s^{-1}}$. We obtained the time constant for an individual mode by measuring the decay of the resonance peak in the power spectrum as a function of time. The faster decay of the oscillations of a charged droplet could be due to the action of the Lorentz force on the oscillating surface charges in the strong magnetic field. A similar effect has been observed in the oscillations of charged helium droplets \cite{whitaker98charged}. The Ohnesorge number, which characterizes the ratio of viscous stresses to surface tension forces, is much less than unity for the sizes of water drops in these experiments. Hence, we expect the viscous damping to have an insignificant effect on the frequencies. For example, taking into account the viscosity of water, we expect the $l=2$ mode of an $a=5$~mm droplet to be just $5\times10^{-4}\%$ lower than that given by Rayleigh's formula\cite{hill10}, \eqn{e1}. Since the time constant of Lorentz-force damping is similar to that of viscous damping, it is reasonable to neglect the action of the Lorentz force on the resonant frequencies as a small effect, too. 

%\section{Conclusion}
In conclusion, we have demonstrated a method to study the dynamics of an electrically charged droplet using diamagnetic levitation. Within the droplet, the force of gravity is compensated at the molecular level, such that the effective gravity acting at all points within the liquid is reduced to within a few percent of $g$. Because of this, diamagnetic levitation allows the levitation of nearly spherical droplets. Using this technique we have confirmed Lord Rayleigh's predicted effect of charge on the normal mode frequencies of a spherical droplet, for the first seven modes. 
% 684

This work was supported by the EPSRC, UK under grants EP/G037647/1 and EP/I004599/1. We are grateful to the technical staff in the School of Physics and Astronomy for constructing the high voltage cavity.  RJAH acknowledges EPSRC for support under a Research Fellowship EP/I004599/1.

%%%%%%%%%%%%%%%%%%%%%%%%%%%%%%%%%%%%%%
% BEFORE SUBMISSION, PASTE THE .bbl FILE HERE,
% AND COMMENT OUT \bibliography{...} COMMAND

%\bibliography{o:/lev}
%Merlin.mbs v4.21 2009-07-09.
%

%%%%%%%%%%%%%%%%%%%%%%%%%%%%%%%%%%%%%%

\end{document}